\documentclass[twocolumn]{aastex631}
\usepackage{amsmath}
\usepackage{booktabs}
\usepackage{multirow}
\usepackage{tikz}
\usepackage{bm}         
\usepackage{pifont}
\makeatletter
\newcommand*{\linktocite}[2]{%
  \hyper@natlinkstart{#1}#2\hyper@natlinkend}
\makeatother

\submitjournal{RNAAS}

\shorttitle{Exoplanet Retrieval Codes}
\shortauthors{MacDonald \& Batalha}

\graphicspath{{./}{Figures/}}

\begin{document}

\title{A Catalogue of Exoplanet Atmospheric Retrieval Codes}

\correspondingauthor{Ryan MacDonald}
\email{ryanjmac@umich.edu}

\author[0000-0003-4816-3469]{Ryan J. MacDonald}
\altaffiliation{NHFP Sagan Fellow}
\affiliation{Department of Astronomy, University of Michigan, 1085 S. University Ave., Ann Arbor, MI 48109, USA}
\author[0000-0003-1240-6844]{Natasha E. Batalha}
\affiliation{NASA Ames Research Center, Moffett Field, CA 94035, USA}

\begin{abstract}

\noindent Exoplanet atmospheric retrieval is a computational technique widely used to infer properties of planetary atmospheres from remote spectroscopic observations. Retrieval codes typically employ Bayesian sampling algorithms or machine learning approaches to explore the range of atmospheric properties (e.g., chemical composition, temperature structure, aerosols) compatible with an observed spectrum. However, despite the wide adoption of exoplanet retrieval techniques, there is currently no systematic summary of exoplanet retrieval codes in the literature. Here, we provide a catalogue of the atmospheric retrieval codes published to date, alongside links to their respective code repositories where available. Our catalogue will be continuously updated via a Zenodo archive.

\end{abstract}


\section{Introduction}
\label{sec:intro}

Spectroscopic observations provide one of our only windows into exoplanet atmospheres. Refining our understanding of atmospheric physics and chemistry ultimately depends on decoding the spectral signatures imprinted in the light from these distant worlds. Given a set of spectral observations of an exoplanet, atmospheric retrieval codes \citep[see][for recent reviews]{Madhusudhan2018,Barstow2020a} invert the spectra to yield probability distributions and detection significances for the underlying atmospheric properties.


While many exoplanet retrieval codes have been developed in the last 15 years, there is no systematic resource summarising the codes reported in the literature. Our goal in this research note is to offer a catalogue summarising the exoplanet retrieval code landscape at the onset of the JWST era. This catalogue also includes links to publicly available retrieval codes, which can facilitate code intercomparisons \citep[e.g.,][]{Barstow2022}.


\section{Exoplanet Retrieval Code Catalogue}
\label{sec:retrieval_codes}

Table~\ref{tab:retrieval_codes} contains our initial exoplanet retrieval code catalogue. At the time of writing (March 2023), we have identified 50 distinct retrieval codes in the exoplanet literature. Of these codes, 39 employ at least one Bayesian sampling technique, whilst 11 codes adopt machine learning algorithms. We sort the codes according to a rough chronology of their first atmospheric retrieval application (on either real or simulated observations) and provide references for the first application of each code to each class of exoplanet spectrum (transmission, emission, reflection, etc.). We apologise for any errors or omissions in the initial catalogue.

We will continuously update this resource by hosting a live version of the catalogue on a \href{https://doi.org/10.5281/zenodo.7675743}{Zenodo repository}. Given the fast-moving nature of the exoplanet field, we encourage authors to contact us with any updates 
or corrections to ensure this catalogue can remain a current and helpful resource for the exoplanet community. 

\startlongtable
\begin{deluxetable*}{lllcl}\label{tab:retrieval_codes}
\tablecaption{Catalogue of Exoplanet Atmospheric Retrieval Codes}
\tabletypesize{\footnotesize}
\tablehead{
\multirow{2}{*}{Code / Authors} & Spectrum & Parameter & Code & \multirow{2}{*}{References} \\
 & \hspace{5pt} Type & Exploration & Link & 
}
\startdata \\[-10pt]
\hspace{-0.2cm} \underline{\textbf{Sampling Based}} & & & & \\[2pt] 
\multirow{2}{*}{Madhusudhan \& Seager} & Transmission & \multirow{2}{*}{Grid, MCMC} & \multirow{2}{*}{---} & \multirow{2}{*}{\citet{Madhusudhan2009}} \\[-2pt]
& Emission & & & \\[2pt]
\multirow{3}{*}{NEMESIS} & Emission & \multirow{3}{*}{OE, NS} & \multirow{3}{*}{\href{https://github.com/nemesiscode/radtrancode}{Link}} & \citet{Lee2012} \\[-2pt]
& Transmission & & & \citet{Barstow2013} \\[-2pt]
& Reflection & & & \citet{Barstow2014} \\[2pt]
\multirow{2}{*}{SCARLET} & Transmission & \multirow{2}{*}{MCMC, NS} & \multirow{2}{*}{---} & \citet{Benneke2012} \\[-2pt]
& Emission & & & \citet{Benneke2019} \\[-2pt]
& Reflection & & & \citet{Wong2020} \\[2pt]
MassSpec & Transmission & MCMC & --- & \citet{deWit2013} \\[2pt]
\multirow{3}{*}{CHIMERA} & Emission & \multirow{3}{*}{OE, MCMC, NS, SC-Grid} & \multirow{3}{*}{\href{https://github.com/mrline/CHIMERA}{Link}} & \citet{Line2013} \\[-2pt]
& Transmission & & & \citet{Swain2014} \\[-2pt]
& Reflection & & & \citet{Piskorz2018} \\[2pt]
\multirow{2}{*}{TauREx} & Transmission & \multirow{2}{*}{MCMC, NS} & \multirow{2}{*}{\href{https://github.com/ucl-exoplanets/TauREx3_public}{Link}} & \citet{Waldmann2015a} \\[-2pt]
& Emission & & & \citet{Waldmann2015b} \\[2pt]
Lupu et al. & Reflection & MCMC, NS & --- & \citet{Lupu2016} \\[2pt]
HELIOS-R & Emission & NS & \href{https://github.com/exoclime/Helios-r2}{Link} & \citet{Lavie2017} \\[2pt]
\multirow{2}{*}{APOLLO} & Transmission & \multirow{2}{*}{MCMC} & \multirow{2}{*}{\href{https://github.com/alexrhowe/APOLLO}{Link}} & \citet{Howe2017} \\[-2pt]
& Emission & & & \citet{Howe2022} \\[2pt]
\multirow{2}{*}{POSEIDON} & Transmission & \multirow{2}{*}{NS} & \multirow{2}{*}{\href{https://github.com/MartianColonist/POSEIDON}{Link}} & \citet{MacDonald2017a} \\[-2pt]
& Emission & & & \citet{Coulombe2023} \\[2pt]
\multirow{2}{*}{ATMO} & Transmission & \multirow{2}{*}{MCMC, NS, SC-Grid} & \multirow{2}{*}{---} & \citet{Wakeford2017} \\[-2pt]
& Emission & & & \citet{Evans2017} \\[2pt]
Brewster & Emission & MCMC, NS & --- & \citet{Burningham2017} \\[2pt]
\multirow{2}{*}{Pyrat Bay} & Transmission & \multirow{2}{*}{MCMC} & \multirow{2}{*}{\href{https://github.com/pcubillos/pyratbay}{Link}} & \citet{Kilpatrick2018} \\[-2pt]
& Emission & & & \citet{Cubillos2021} \\[2pt]
HyDRA & Emission & NS & --- & \citet{Gandhi2018} \\[2pt]
\multirow{3}{*}{PSG} & Reflection & \multirow{3}{*}{OE, NS} & \multirow{3}{*}{\href{https://github.com/nasapsg}{Link}} & \multirow{3}{*}{\citet{Villanueva2018}} \\[-2pt]
& Emission & & & \\[-2pt]
& Transmission & & & \\[2pt]
AURA & Transmission & NS & --- & \citet{Pinhas2018} \\[2pt]
exoretrievals & Transmission & NS & --- & \citet{Espinoza2019} \\[2pt]
Brogi \& Line & Emission & NS & \href{https://www.dropbox.com/sh/0cxfolfmrs8ip37/AABZYoHr8nuRlHJG84dArX4ea?dl=0}{Link} & \citet{Brogi2019} \\[2pt]
\multirow{2}{*}{PLATON} & Transmission & \multirow{2}{*}{NS} & \multirow{2}{*}{\href{https://github.com/ideasrule/platon}{Link}} & \citet{Zhang2019} \\[-2pt]
& Emission & & & \citet{Zhang_M_2020} \\[2pt]
\multirow{3}{*}{petitRADTRANS} & Transmission & \multirow{3}{*}{MCMC, NS} & \multirow{3}{*}{\href{https://gitlab.com/mauricemolli/petitRADTRANS}{Link}} & \citet{Molliere2019} \\[-2pt]
& Emission & & & \citet{Molliere2019} \\[-2pt]
& Reflection & & & \citet{Alei2022} \\[2pt]
MERC & Transmission & NS & --- & \citet{Seidel2020} \\[2pt]
species & Emission & MCMC, NS, SC-Grid & \href{https://github.com/tomasstolker/species}{Link} & \citet{Stolker2020} \\[2pt]
Gibson et al. & Transmission & MCMC & --- & \citet{Gibson2020} \\[2pt]
ExoReL$^{\mathcal{R}}$ & Reflection & NS & --- & \citet{Damiano2020} \\[2pt]
Alfnoor & Transmission & NS & --- & \citet{Changeat2020} \\[2pt]
PETRA & Transmission & MCMC, SC-Grid & --- & \citet{Lothringer2020a} \\[2pt]
METIS & Transmission & MCMC & --- & \citet{Lacy2020a} \\[2pt]
Carrión-González et al. & Reflection & MCMC & --- & \citet{Carrion-Gonzalez2020} \\[2pt]
\multirow{2}{*}{ARCiS} & Transmission & \multirow{2}{*}{NS} & \multirow{2}{*}{---} & \citet{Min2020} \\[-2pt]
& Emission & & & \citet{Chubb2022} \\[2pt]
\multirow{3}{*}{PICASO} & Reflection & \multirow{3}{*}{NS, SC-Grid} & \multirow{3}{*}{\href{https://github.com/natashabatalha/picaso}{Link}} & \citet{Mukherjee2021} \\[-2pt]
& Emission & & & \citet{Miles2022} \\[-2pt]
& Transmission & & & \citet{Batalha2023} \\[2pt]
Cerberus & Transmission & MCMC & --- & \citet{Swain2021} \\[2pt]
Aurora & Transmission & NS & --- & \citet{Welbanks2021} \\[2pt]
\multirow{2}{*}{BART} & Transmission & \multirow{2}{*}{MCMC} & \multirow{2}{*}{\href{https://github.com/ExOSPORTS/BART/}{Link}} & \multirow{2}{*}{\citet{Harrington2022}} \\[-2pt]
& Emission & & & \\[2pt]
ExoJAX & Emission & MCMC & \href{https://github.com/HajimeKawahara/exojax}{Link} & \citet{Kawahara2022} \\[2pt]
ThERESA & Eclipse Mapping & MCMC & \href{https://github.com/natashabatalha/picaso}{Link} & \citet{Challener2022} \\[2pt]
p-winds & Transmission & MCMC & \href{https://github.com/ladsantos/p-winds}{Link} & \citet{DosSantos2022} \\[2pt]
smarter & Transmission & NS & --- & \citet{Lustig-Yaeger2022} \\[2pt]
tierra & Transmission & MCMC & \href{https://github.com/disruptiveplanets/tierra}{Link} & \citet{Niraula2022} \\[2pt]
\multirow{3}{*}{rfast} & Reflection & \multirow{3}{*}{MCMC} & \multirow{3}{*}{\href{https://github.com/hablabx/rfast}{Link}} & \multirow{3}{*}{\citet{Robinson2023}} \\[-2pt]
& Emission & & & \\[-2pt]
& Transmission & & & \\[0pt]
\midrule
\hspace{-0.2cm} \underline{\textbf{Machine Learning}} & & & & \\[1pt]
HELA & Transmission & RF & \href{https://github.com/exoclime/HELA}{Link} & \citet{Marquez-Neila2018} \\[1.5pt]
ExoGAN & Transmission & NN & \href{https://github.com/ucl-exoplanets/ExoGAN_public}{Link} & \citet{Zingales2018} \\[1.5pt]
\multirow{2}{*}{INARA} & Reflection & \multirow{2}{*}{NN} & \multirow{2}{*}{\href{https://gitlab.com/frontierdevelopmentlab/astrobiology/inara}{Link}} & \multirow{2}{*}{\citet{Soboczenski2018}} \\[-2pt]
& Emission & & & \\[1.5pt]
plan-net & Transmission & NN & \href{https://github.com/exoml/plan-net}{Link} & \citet{Cobb2019} \\[1.5pt]
Fisher et al. & Transmission & RF & --- & \citet{Fisher2020} \\[1.5pt]
Johnsen \& Marley & Reflection & MLP & \href{https://github.com/WreckItTim/MLP-Estimating-Exoplanet-Parameters}{Link} & \citet{Johnsen2020} \\[1.5pt]
Nixon \& Madhusudhan & Transmission & RF & --- & \citet{Nixon2020} \\[1.5pt]
MARGE+HOMER & Emission & NN+MCMC & \href{https://github.com/exosports/homer}{Link} & \citet{Himes2022} \\[1.5pt]
exoCNN & Transmission & NN & \href{https://gitlab.astro.rug.nl/ardevol/exocnn}{Link} & \citet{ArdevolMartinez2022} \\[1.5pt]
VI-retrieval & Transmission & NN+VI & --- & \citet{Yip2022} \\[1.5pt]
Vasist et al. & Emission & NN+VI & --- & \citet{Vasist2023} \\[1.5pt]
\enddata
\tablecomments{Retrieval codes are ordered according to their first published exoplanet retrieval application (on either real or simulated spectroscopic data) for each category of exoplanet spectra. For years when multiple codes were published, the order is determined by the date of first arXiv submission. `Emission' refers to secondary eclipse emission spectra, phase-resolved emission spectra, or direct thermal emission from a self-luminous object (i.e. directly imaged exoplanets or brown dwarfs). Parameter exploration acronyms are as follows: Optimal Estimation (OE); Markov Chain Monte Carlo (MCMC); Nested Sampling (NS); Self-Consistent Grid (SC-Grid); Random Forest (RF); Neural Network (NN); Multilayer Perceptron (MLP); and Variational Inference (VI). Publicly available retrieval codes, at the time of this publication, have links to the code repository.}
\end{deluxetable*}

\newpage

\begin{acknowledgments}
R.J.M. thanks Eileen Gonzales, Niall Whiteford, Nikole Lewis, and Joanna Barstow for helpful discussions while assembling this catalogue.
\end{acknowledgments}

\newpage

\bibliography{retrieval_catalogue}{}
\bibliographystyle{aasjournal}

\end{document}